\documentstyle[epsfig,prc,aps,preprint,tighten]{revtex}
                      
\newcommand{\fslash}{\not \!}

\input{prepictex} \input{pictex} \input{postpictex}
\begin{document}
\draft

\title{Role of $\Delta$ exchange for $p \bar p$ annihilation into two-pion
and three-pion channels}

\author{M. Betz$^{\rm \, a \, }$\thanks{Email: betz@if.ufrgs.br}, 
E.\ A.\ Veit$^{\rm \, a \, }$\thanks{Email: eav@if.ufrgs.br},
and J. Haidenbauer$^{\rm \, b \, }$\thanks{Email: j.haidenbauer@fz-juelich.de}}

\address{
$^{\rm a}$Instituto de F\'{\i}sica - Universidade Federal do
Rio Grande do Sul  \\
Cx. P. 15051; CEP 91501-970,  Porto Alegre, Brazil
\\
$^{\rm b}$Institut f\"ur Kernphysik, Forschungszentrum
J\"ulich GmbH,
\\ D--52425 J\"ulich, Germany
}

\maketitle

\begin{abstract}
\mbox{$p\bar{p}$} annihilation into two pions and three pions is studied in
a baryon-exchange model. Annihilation diagrams involving nucleon as well as 
$\Delta$-resonance exchanges are included consistently in the two- and
three-pion channels. Effects from the initial-state interaction are 
fully taken into account. 
A comparison of the influence of the $\Delta$ exchange on the considered
annihilation channels is made and reveals that its importance for 
three-pion annihilation is strongly reduced as compared to two-pion
annihilation. It is found that annihilation into three uncorrelated pions
can yield up to 10\% of the total experimentally observed three-pion 
annihilation cross section.
\end{abstract}



\section{Introduction}

Nucleon-antinucleon \mbox{($N \bar N$)} annihilation has attracted a great
deal of interest over the past two decades~\cite{Dover,Amslr}.
Although recently the main emphasis has often been put on meson
spectroscopy and specifically the identification of exotic mesonic
states~\cite{Ams00}, one should not forget that this process also
offers a rich field in which various aspects of quark-gluon dynamics and/or
hadron dynamics can be tested.
Indeed the development of a microscopic
model which describes the \mbox{$N\bar N$} interaction and, at the same time,
can account also for all phenomena associated with the \mbox{$N\bar N$}
annihilation into two, three, ..., mesons is a rather challenging task
for any theorist and has so far not been achieved.
With regard to \mbox{$N\bar N$} annihilation,
models based on quark degrees of freedom were the first ones
to be utilized in attempts to obtain a quantitative description of
annihilation into two and three mesons.
Their use was prompted by the expectation that annihilation processes
might be a good place for detecting explicit quark-gluon effects
since relatively short interaction ranges are involved.
The pioneering
works stem from Maruyama and Ueda~\cite{Ueda} and Green and
Niskanen~\cite{Green}, followed by an impressive series of studies
carried out by the T\"ubingen group~\cite{Tueb}.
(Cf. also Ref.~\cite{Dover} for a comprehensive review of quark-model
studies of {\mbox{$N\bar N$} annihilation.)

Investigations of \mbox{$N\bar N$} annihilation relying on the more traditional
meson-baryon picture, where the annihilation process is described by
baryon-exchange diagrams, were initiated by Moussallam not long after
the first works within the quark model had appeared~\cite{Mouss}.
Subsequently, the J\"ulich group carried out several studies based on
this approach~\cite{Hipp91,Mull95} and more recently also
Yan and Tegen~\cite{Yan96,Yan99}. Those works indicate that the conventional
hadronic concept for describing \mbox{$N\bar N$} annhilation is capable of
producing results that are at least of the same quality as
obtained from quark-gluon models.
However, all those studies concentrated on two-meson annihilation
channels only. This is certainly an unsatisfying situation because
it would be interesting to see whether the baryon-exchange picture of
\mbox{$N\bar N$} annihilation works similarly well for three-meson decay
as it does for the two-meson channels. Furthermore, we wish to
recall that the latest model by the J\"ulich group~\cite{Mull95}
constitutes already an essential step in achieving a unified description 
of \mbox{$N\bar N$}
scattering and annihilation. In this model the elastic and annihilation
parts of the \mbox{$N\bar N$} interaction are derived in a consistent framework
and, in addition, the transitions to two-meson channels,
\mbox{$N\bar N \rightarrow M_1M_2$}, and the contributions of these two-meson
channels to the total \mbox{$N\bar N$} annihilation are described consistently
as well. Thus, it is important, but also challenging, to go a step
further and include the three-meson decay channels
explicitly as well in this model.

In a series of recent works~\cite{us97,usrio,usfloripa,uscaragua},
we have used the baryon-exchange model to investigate proton-antiproton
annihilation into three uncorrelated pions. By ``uncorrelated'', we mean that these
pions are not the decay products of an intermediate heavy-meson resonance. Of course,
annihilation into three pions is known to be dominated by the formation and
decay of intermediate states made up of a pion and a heavy meson, and such processes
have already been studied in the baryon-exchange model. The process at hand only
constitutes a relatively small background. Yet such background contributions have
been found significant in phenomenological analyses of \mbox{$p \bar{p}$}
annihilation into \mbox{$\pi^+\pi^- \pi^0$} at rest~\cite{Foster,ASTERIX}
and it can be expected that they need to be also taken into account in precise
analyses of annihilation into three neutral pions, which has recently been the
object of considerable interest~\cite{Amsler,Abele,Anisovich1999452}.

Our investigation of annihilation into three uncorrelated pions
was done in a distorted-wave Born approximation (DWBA), employing the
most recent \mbox{$N \bar N$} model of the J\"ulich group~\cite{Mull95} for the
initial-state interaction, and guided by the same principles applied in
that reference for the calculation of annihilation into two-meson
channels. The work was initiated
with a study of the \mbox{$p \bar{p} \rightarrow \pi^+\pi^- \pi^0$}
process~\cite{us97}, in a simplified model, in which only nucleon exchange was
included explicitly in the annihilation amplitude.
The effect of $\Delta$ exchange -- which is known to produce a
significant enhancement of the cross-section for annihilation into
two pions~\cite{Mouss,Yan96} -- was taken into account in an approximate
phenomenological fashion, namely through a readjustment of the cutoff at the
\mbox{$N N \pi$} vertex.
In a subsequent paper~\cite{usrio}, a first step in assessing the
shortcomings of this simplification was made by comparing
the pertinent results with those obtained through
explicit inclusion of the amplitudes involving the exchange of one $N$ and
one $\Delta$. Very recently, a further step
was taken with the inclusion of the annihilation amplitude generated by
double $\Delta$ exchange~\cite{usfloripa,uscaragua}.

Despite those achievements we have to concede, however, that there is
still an inconsistency in our calculations so far.
The nucleon-antinucleon interaction developed
by the J\"ulich group is derived in a time-ordered formalism and
specifically also the amplitude for annihilation into two pions is
calculated in time-ordered perturbation theory.
In contrast, in order to avoid the proliferation of diagrams, the three-pion
annihilation amplitude is calculated in Feynman-type perturbation theory.
This procedure implies that the treatment of the off-shell behaviour of the
three-pion annihilation amplitude is different from that used in the
J\"ulich model of annihilation into two pions~\cite{Mull95}.
This shortcoming has been previously ignored. However, it is of relevance now
that we are able to calculate the two-pion and three-pion annihilation channels
involving $N$ as well as $\Delta$ exchange in a consistent way. Therefore, we
decided to treat the two-pion annihilation amplitude
in the same fashion as the three-pion one, i.e.\ in Feynman-type perturbation
theory. Clearly, in such an approach \mbox{$NN\pi$} and \mbox{$N\Delta\pi$} vertex
parameters taken from the work of the J\"ulich group will no longer reproduce
the phenomenology of two-pion annihilation and a refitting is
therefore necessary. This means, in turn, that full consistency with the
J\"ulich \mbox{$N\bar N$} model, which we continue to use as the initial-state
interaction, will be lost. But this is excusable to a certain extent because the
annihilation channel in question, \mbox{$N\bar N \rightarrow 2\pi$}, yields only a tiny
contribution to the total \mbox{$N\bar N$} annihilation cross section~\cite{Mull95}.
On the other hand, the \mbox{$NN\pi$} and \mbox{$N\Delta\pi$} vertex parameters play a
crucial role for the \mbox{$N\bar N \rightarrow 3\pi$} cross section, as we will see
and explore in the present paper, and therefore it is rather important to
constrain them by the requirement of consistency between the two- and three-pion
annihilation channels.

In this paper we present results of a combined study of
proton-antiproton annihilation into two and three pions in a baryon-exchange
model. We start out from the two-pion annihilation channel. We use
available experimental data on \mbox{$p\bar p \rightarrow \pi^+\pi^-$}
to determine the free parameters of our model, i.e.
the cutoff masses in the form factors at the \mbox{$NN\pi$} and \mbox{$N\Delta\pi$} vertices.
The effects of different choices for the analytical form of those vertex
form factors are explored as well. We then turn to three-pion annihilation
and discuss the relative importance of \mbox{$NN$}, \mbox{$N\Delta$} and \mbox{$\Delta\Delta$}
exchanges. 

In Sec.~\ref{sec:the model} we provide some details of our model. In particular, we specify 
the ingredients used for evaluating the $N$- and $\Delta$-exchange diagrams
for \mbox{$N\bar N$} annihilation into two and three pions, i.e. the baryon-baryon-meson
Lagrangians and the corresponding vertex form factors and coupling constants. 
Furthermore we give a short description of the \mbox{$N\bar N$} model that is employed
for the initial-state interaction and we outline how the amplitudes
for \mbox{$N\bar N$} annihilation into two and three pions are determined in distorted-wave
Born approximation.
Our results are presented and discussed in Sec.~\ref{sec:results}. First we consider the
reaction \mbox{$p\bar p \rightarrow \pi^+\pi^-$} which is used for fixing the free
parameters of our model. Subsequently, we examine the reaction 
\mbox{$p\bar p \rightarrow \pi^+\pi^-\pi^0$} as well as annihilation channels 
involving only neutral pions (\mbox{$p\bar p \rightarrow \pi^0\pi^0$},
\mbox{$p\bar p \rightarrow \pi^0\pi^0\pi^0$}). We show results for total and 
differential cross sections and also for branching ratios from specific
initial \mbox{$N\bar N$} states. The paper ends with some concluding remarks.

\section{THE MODEL}
\label{sec:the model}
\subsection{Distorted-wave Born approximation}

The basis of the present work is the J\"ulich model for \mbox{$N\bar N$} scattering
and annihilation~\cite{Mull95}. In this model heavier mesons are
treated as stable particles in the (successful) description of \mbox{$N\bar{N}$}
annihilation into two mesons. In accordance with this approach, the uncorrelated three-pion
channel is here considered separately from the channels in which a pion and a heavy
meson, which can decay into two pions, are formed. When calculating
the total cross section for annihilation into three pions, the contributions from
these channels are added incoherently. Neglecting interferences between
the various channels seems acceptable for the aim of this work, which is
to perform an exploratory study about the
relevance of the uncorrelated  three-pion channel for the annihilation cross section.

The general procedure is to start from the Born transition amplitude
\mbox{$V^{N\bar{N}\rightarrow n\pi}$}
for annihilation into $n$ (= 2 and 3) pions and include the initial-state
interaction in distorted-wave Born approximation, so that the annihilation
amplitude \mbox{$T^{N\bar N \rightarrow n\pi}$} is given by

\begin{equation}
T^{N \bar{N} \rightarrow n\pi} = V^{N \bar{N} \rightarrow n\pi} +
V^{N \bar{N} \rightarrow n\pi}
G^{N \bar{N} \rightarrow N \bar{N}}
T^{N \bar{N} \rightarrow N \bar{N}},
\label{ampl}
\end{equation}
where \mbox{$G^{N \bar{N} \rightarrow N \bar{N}}$} is the propagator for the
\mbox{$N \bar{N}$} pair. The \mbox{$N \bar{N}$} scattering
amplitude \mbox{$T^{N \bar{N} \rightarrow N \bar{N}}$} is obtained from the
solution of a Lippmann-Schwinger equation~\cite{Mull95}

\begin{equation}
T^{N \bar{N} \rightarrow N \bar{N}} = V^{N \bar{N} \rightarrow N
\bar{N}}
+ V^{N \bar{N} \rightarrow N \bar{N}} G^{N \bar{N} \rightarrow N
\bar{N}}
T^{N \bar{N} \rightarrow N \bar{N}}.
\label{TNN}
\end{equation}

In the work of the J\"ulich group the time-ordered formalism is invoked to cast
Eqs.~(\ref{ampl}) and (\ref{TNN}) in tractable (three-dimensional) form.
The \mbox{$N\bar N$} interaction \mbox{$V^{N\bar N \rightarrow N\bar N}$} and, in particular,
the Born transition amplitudes for annihilation into two mesons are likewise
calculated within time-ordered perturbation theory (TOPT).
Here however, in order to avoid the evaluation of the numerous graphs
that occur within TOPT for the Born transition amplitude \mbox{$V^{N \bar{N}
\rightarrow 3\pi}$} for annihilation into three pions,
we prefer to employ Feynman diagrams. We stress that these two
procedures imply different off-shell extrapolations of the annihilation amplitudes and
are therefore not equivalent. Since we intend to use annihilation into two pions
for the determination of the free parameters in our model, we need to maintain
consistency between our treatments of annihilation into two and three pions.
Therefore, we
will use the Feynman prescription also to calculate the Born transition amplitude
\mbox{$V^{N \bar{N} \rightarrow 2\pi}$} for annihilation into two pions. Only the
initial-state scattering amplitude \mbox{$T^{N \bar{N} \rightarrow N \bar{N}}$}
will be still computed in time-ordered formalism.

In the case of annihilation into two pions, the amplitudes may be expanded in partial
waves and the angular distribution and integrated cross section calculated in
standard fashion. In the case of annihilation into three pions, the Monte Carlo method is used
to perform the final phase-space integration.

\subsection{Annihilation amplitudes}
\label{sec:annihilation amplitudes}

In accordance with the J\"ulich \mbox{$N\bar N$} model we assume that the 
dynamics of annihilation into pions is
mediated by nucleon and $\Delta$ exchanges. Hence, the Born transition amplitudes
\mbox{$V^{N \bar{N} \rightarrow n\pi}$} (\mbox{$n=2,3$}) are given by the sums 
of the Feynman tree
diagrams involving the \mbox{$N N \pi$}, \mbox{$\Delta N \pi$} and 
\mbox{$\Delta \Delta \pi$} vertices,
shown in Fig.~\ref{fig1} and Fig.~\ref{fig2}. The complete amplitudes are obtained
by summing such diagrams over all permutations of the final-state pions.

The contributions of the various diagrams are evaluated using standard 
interaction Lagrangians, namely
\begin{eqnarray}
{\cal L}_{N N \pi} & = & \frac{f_{NN \pi}}{m_\pi} \bar{\psi}_N
\gamma^5 \vec{\tau} \cdot \fslash  \partial \vec{\phi} \, \psi_N \; , \\
{\cal L}_{\Delta N \pi} & = & \frac{f_{\Delta N \pi}}{m_\pi}
\bar{\psi}_\Delta^{\, \mu}
 \vec{T} \cdot \partial_\mu \vec{\phi} \, \psi_N + h. c. \; , \\
{\cal L}_{\Delta \Delta \pi} & = & - \frac{f_{\Delta \Delta \pi}}{m_\pi}
 \bar{\psi}_{\Delta \mu} \, \gamma^5 \vec{I}
 \cdot \fslash  \partial \vec{\phi} \, \psi_\Delta^\mu  \; ,
\end{eqnarray}
where $\vec{I}$ is the $\Delta$ isospin operator and $\vec{T}$ the
\mbox{$N \rightarrow \Delta$} transition isospin operator (we use the normalization
conventions of Ref.~\cite{Schu} for these operators).

The corresponding vertex factors in Feynman diagrams are:
\begin{eqnarray}
{\cal V}_{N N \pi} & = & - \frac{f_{NN \pi}}{m_\pi} \gamma^5 \tau^i \fslash q   \; , \\
{\cal V}_{\Delta N \pi} & = & -  \frac{f_{\Delta N \pi}}{m_\pi} q_\mu
 T^i  + h. c. \; , \\
{\cal V}_{\Delta \Delta \pi} & = & \frac{f_{\Delta \Delta \pi}}{m_\pi} g_{\mu \nu}
 \gamma^5 I^i \fslash q \; ,
\end{eqnarray}
where $q_\mu$ is the pion four-momentum and the index $i$ specifies the pion isospin.
The nucleon propagator takes the standard form
\begin{equation}
i G_N(p) = i \frac{\fslash p+m_N}{p^2-m_N^2} \; .
\label{nucleon propagator}
\end{equation}
For the $\Delta$ propagator, we adopt~\cite{Ben}
\begin{equation}
i G_{\Delta}^{\mu \nu}(p) = - i \frac{\fslash p+ m_{\Delta}}
{p^2-m_{\Delta}^2} \Theta^{\mu \nu} (p) \; ,
\label{delta propagator}
\end{equation}
with
\begin{equation}
\Theta^{\mu \nu} (p) \equiv  g^{\mu \nu} - \frac{\gamma^\mu \gamma^\nu}
{3} - \frac{2 p^\mu p^\nu}{3 m_\Delta^2} +\frac{p^\mu \gamma^\nu -
p^\nu \gamma^\mu}{3 m_\Delta} \; .
\label{Theta}
\end{equation}

Two of the coupling constants appearing in the above vertices are already used in the
Bonn potential~\cite{Machleidt} and J\"ulich \mbox{$N\bar{N}$} model~\cite{Mull95},
namely \mbox{$f_{NN \pi}^2/4 \pi$ = 0.0778} and \mbox{$f_{\Delta N \pi}^2/4 \pi$ = 0.224}. 
For the \mbox{$\Delta\Delta\pi$} coupling, we shall rely
on the \mbox{$SU(2) \times SU(2)$} quark--model
relation~\cite{JJdS}, \mbox{$f_{\Delta \Delta \pi}
= \frac{9}{5} f_{NN \pi}$}. This gives \mbox{$f_{\Delta \Delta \pi}^2/4 \pi$ = 0.252}.

Form factors must be included in order to regularize the calculation and
to take into account the extended hadron structure. We know from our experience
with other hadronic reactions that the results might depend sensitively on those
form factors, in particular when loop integrations like in Eq.~(\ref{ampl}) are involved.
Therefore, in order
to investigate the sensitivity of our model to the details of these form factors, we
employ three different parametrizations in the present study. In two of these we
follow the conventional assumption that the vertex form factor depends only on the
momentum of the exchanged particle, i.e.\ the particle which is off-shell in the
Born diagram. For annihilation into three pions, the inner vertex contains the product
of two such form factors since there are two off-shell particles attached to this vertex.
Explicitly, we employ the functions

\begin{equation}
{\cal F}_M(p) = \frac{\Lambda^2_{\alpha  X\pi}-M^2_X}{\Lambda^2_{\alpha X \pi}- p^2}\;
\label{mono}
\end{equation}
and
\begin{equation}
{\cal F}_G (p) = exp{\Big  [
- \frac{(p^2 - M_X^ 2)^2}{\Lambda^4_{\alpha X\pi}}
\Big ]} \; .
\label{gaus}
\end{equation}

In these expressions, $X$ stands for the type ($N$ or $\Delta$) of the exchanged off-shell
particle, $M_X$ for its mass and $p$ for its four-momentum; $\alpha$ denotes the type of the
other baryon present at the vertex and \mbox{$\Lambda_{\alpha X\pi}$} represents the cutoff mass
corresponding to the \mbox{$\alpha X\pi$} vertex. In principle, four independent cutoff masses have
to be specified: one for the \mbox{$NN \pi$} vertex, characterizing an off-shell nucleon,
two for the \mbox{$\Delta N \pi$} vertex, characterizing an off-shell nucleon
or an off-shell $\Delta$, respectively, and one for the \mbox{$\Delta \Delta \pi$} vertex,
characterizing an off-shell $\Delta$. In order to reduce the number of free parameters,
we assume \mbox{$\Lambda_{N\Delta\pi} = \Lambda_{\Delta N\pi}=\Lambda_{\Delta\Delta\pi} \equiv
\Lambda_\Delta$}. The procedure for fixing the values of the remaining independent cutoff
parameters (\mbox{$\Lambda_{NN\pi} \equiv \Lambda_N$} and $\Lambda_\Delta$) will be
discussed in Sec.~\ref{sec:results}. In the following we will refer to those form factors as
monopole [Eq.~(\ref{mono})] and Gaussian [Eq.~(\ref{gaus})], respectively.

The third type of regularization we consider attributes a damping factor to each
line in the Born diagrams, including the external legs. Specifically we use
a parametrization introduced by B.\ Pearce~\cite{Pearce} and later by C.\
Sch\"utz~\cite{Schutz94} in their studies of the \mbox{$\pi N$} system
which is given by

\begin{equation}
{\cal F}_P(p) =  \frac{\Lambda^4_X}{\Lambda^4_X+(p^2 - M_X^2)^2}
 \; ,
\label{pear}
\end{equation}
with $\Lambda_X$ the cutoff mass associated with a baryon line of type $X$.
Note that, in principle, a similar factor should be applied to the pion lines
as well. But since in our DWBA calculation the pions are always on their mass
shell this factor will be identical to 1 and, therefore, can be omitted.
In the following we will refer to this choice as the Pearce form factor.

In order to combine the Feynman annihilation amplitudes with the initial-state
distorted wave, it is necessary to specify a prescription for the energy components
of the four-momenta of their external $N$ and $\bar{N}$ legs. We set both equal to
their on-energy-shell value (i.e., half the total available energy in the center-of-mass
frame).

\subsection{Initial-state interaction}

The \mbox{$N \bar{N}$} interaction \mbox{$V^{N \bar{N} \rightarrow N \bar{N}}$}
used to obtain the initial-state distorted wave is that developed in Ref.~\cite{Mull95},
without modification. For completeness, we summarize here its main features.

The interaction is made up of an elastic and an annihilation part:

\begin{equation}
V^{N \bar{N} \rightarrow N \bar{N}} = V_{el} + V_{ann}.
\end{equation}

The elastic interaction is obtained through a G-parity transformation of the
 full Bonn \mbox{$NN$} potential~\cite{Machleidt}, corresponding to the diagrams shown in
Fig.~\ref{fig3}.

The annihilation interaction consists of a microscopic and a phenomenological
 piece:
\begin{equation}
V_{ann} =    \sum_{ij} V^{M_i M_j \rightarrow N \bar{N}}
G^{M_i M_j} V^{N \bar{N}
\rightarrow M_i M_j}  + V_{opt}.
\end{equation}
The microscopic component is the sum of box diagrams with two-meson intermediate states
resulting from all possible combinations of
$\pi$, $\eta$, $\rho$, $\omega$, $a_0$, $f_0$, $a_1$, $f_1$, $a_2$, $f_2$,
$K$ and $K^*$ mesons (Fig. \ref{fig4}a). The transition potentials 
\mbox{$V^{N \bar{N} \rightarrow M_i M_j}$} are given by the baryon-exchange diagrams 
presented in Fig.~\ref{fig5}. The coupling constants and cutoff parameters
at the vertices of these transition potentials are quoted in Table I of Ref.~\cite{Mull95}.
Note that in Ref.~\cite{Mull95} these transition potentials are employed also 
for the calculation of the amplitudes for annihilation into a pion and a heavy 
meson $M$. Those amplitudes are obtained from 
equations analogous to Eq.~(\ref{ampl}), with \mbox{$V^{N \bar{N} \rightarrow \pi M}$} 
as the Born term. The corresponding contributions to the cross sections for annihilation 
into three pions will be used here without modification.

The phenomenological optical potential (Fig.~\ref{fig4}b) simulates the effect of
contributions from annihilation into more than two mesons, and
is parametrized in coordinate space as
\begin{equation}
V_{opt} = - i W \exp{(-\frac{r^2}{2r_0^2})},
\end{equation}
with \mbox{$W = 1\, {\rm GeV}$} and \mbox{$r_0 = 0.4 \, {\rm fm}$}. These values have
been obtained~\cite{Mull95} through an overall fit to \mbox{$N \bar{N}$}
integrated cross-section data.

\section{RESULTS AND DISCUSSION}
\label{sec:results}

As mentioned in the Introduction, in a previous publication~\cite{us97} aimed at acquiring
a first idea of the relevance of the uncorrelated channel to the
\mbox{$p\bar{p}\rightarrow \pi^+\pi^-\pi^0$} cross section, $\Delta$ exchange
was not included in the transition amplitude. This, of course, is at variance with
the dynamics included in the treatment of the \mbox{$N\bar{N}\rightarrow 2\pi$}
 amplitude, where $\Delta$ exchange is taken into account.
The argument invoked was that if one considers only the (dominant) charged two-pion
channel, one finds that the effect of $\Delta$ exchange can be described
phenomenologically by using an effective value for the cutoff at the \mbox{$NN\pi$}
vertex. It turned out that, in this rather crude treatment, the
uncorrelated channel  adds a 10\% contribution to the total \mbox{$p\bar{p}$} annihilation cross
section into three pions, motivating a more systematic study. In subsequent studies, the
exchange of one $\Delta$~\cite{usrio}, and of two $\Delta$'s~\cite{usfloripa} has been explored.
While this unified the dynamics of annihilation into two and three pions,
complete consistency was not achieved yet, since the two-pion channel was treated
with TOPT~\cite{Mull95} while for the three-pion channel, the Feynman
prescription was used~\cite{usrio,usfloripa}. More recently~\cite{uscaragua}, first
results of a calculation in which both two-pion and three-pion channels are
treated within the Feynman prescription were presented. Here we explore more fully
the predictions of this model. Specifically we analyze the contributions
of the uncorrelated three-pion channel not only to the charged but
also to the neutral annihilation cross sections, and we show predictions for
various branching ratios as well.
We also investigate the effect of
various choices of the vertex form factors on our results and we explore
the importance of the contributions coming from the $\Delta$-exchange diagrams.

\subsection{\mbox{$p\bar{p}\rightarrow \pi^+\pi^-$} cross section}
To get insight into the influence of the vertex form factors
on the results for \mbox{$p\bar p \rightarrow 3\pi$} we accomplished fits to the
reaction \mbox{$p\bar p \rightarrow \pi^+\pi^-$}
using each of the three parametrizations introduced in Sec.~\ref{sec:annihilation amplitudes}.
Since it turned out that the presently available data do not allow to
determine the relative magnitudes of the $N$ and $\Delta$ exchange
contributions unambiguously, we prepared two sets of models, one with
$\Lambda_N$ larger than $\Lambda_\Delta$ and the other with $\Lambda_N$ smaller
than $\Lambda_\Delta$.
Thus, we have in total six parameter sets, all determined in such a way as to
provide an adequate description of the total and differential
\mbox{$p\bar{p}\rightarrow \pi^+\pi^-$} cross sections.
The resulting values for the cutoff masses are compiled in
Table~\ref{table1}. The parameter set~A corresponds to form factors which
suppress $\Delta$ exchange compared to nucleon exchange. The parameter set B
corresponds to a quite strong $\Delta$ exchange, hopefully providing some sort of
upper bound for the $\Delta$ contribution when used to describe annihilation into
three uncorrelated pions later on.

The results for the \mbox{$p\bar{p}\rightarrow \pi^+\pi^-$} total cross sections 
are compared to the data \cite{Sai,Sugimoto,Bardin}
in Fig.~\ref{fig6}. Considering the rather large experimental errors at
low energy, all parameter sets can be said to provide an adequate description.
The relative importance of $N$ and $\Delta$ exchanges is illustrated in 
Table~\ref{table2},
which shows their separate contributions to the cross section at two sample energies.
For the parameter set~A the contribution from $\Delta$ exchange is indeed
not very important. Turning it off reduces the cross sections by -- at most, 
depending on the energy -- 30\% (Pearce form factor) to  10\% 
(monopole form factor).  For the Gaussian form
factor, the reduction is by about 20\%. For the parameter set B, the $\Delta$-exchange
contribution dominates over the $N$ contribution, although, because of the constructive
interference between them, both contributions are important in building up the cross section.
One should note that the relative importance of $N$ and $\Delta$ exchanges results in general
from a balance between two competing effects. On the one hand, the larger coupling constant
at the \mbox{$\Delta N \pi$} vertex favors $\Delta$ exchange over $N$ exchange, but on the other
hand, since the relevant kinematical region for the exchanged particle is space-like, it
is somewhat further away from the on-shell point for the $\Delta$ than for the nucleon.
This leads to a comparatively larger propagator denominator and a stronger damping due to form 
factors for the $\Delta$.  

The angular distributions are shown in Fig.~\ref{fig7} for two incident momenta.
The quality of the
results for other measured incident momenta is similar to that of these two sample cases. For
forward scattering, the differential cross section looks reasonable for all parameter sets.
For larger incident momenta, a stronger $\Delta$ contribution tends to produce a bending down
of the cross section at small angles (see the case \mbox{$p_{lab} = 679$~MeV/c} 
with parameter set B). This imposes an upper limit on the amount of $\Delta$ contribution which is acceptable.
Let us note, however, that the data at somewhat higher energies do indeed show such a
bending down in the forward direction~\cite{Hasan}.
With respect to backward scattering our models fare rather poorly. But this is
expected and therefore not a reason of concern. It is known that
 final-state interactions, not considered
here, play an important role for backward angles and do improve the results
significantly~\cite{Mull92}. Also, it has been argued~\cite{Yan96}
that a tensor coupling for the \mbox{$N\Delta\pi$} vertex would enhance the
differential cross section around $100^\circ$ for \mbox{$p_{lab} > 680$~MeV/c}.

\subsection{\mbox{$p\bar{p} \rightarrow \pi^+\pi^-\pi^0$} annihilation}
\subsubsection{Cross section}
\label{sec:3 charged cc}
The \mbox{$p\bar{p}$} annihilation into three pions has contributions from three uncorrelated
pions in addition to some two-meson channels, involving a heavier meson  ($\rho$, $f_o$ and
$f_2$), which decays subsequently into two pions with some branching ratio (100\%, 78\% and
85\%, respectively). The two-meson channels have already been studied with the J\"ulich model
by Mull et al.~\cite{Mull95}. Here we concentrate on the uncorrelated three-pion contribution
to the \mbox{$p\bar{p}\rightarrow \pi^+\pi^-\pi^0$} cross section.

Our results are shown in Table~\ref{table3} for the six parameter sets of 
Table~\ref{table1}, at two sample energies.
For parameter set~A most of the cross section comes from $N$ exchange and the contributions
from one- and two-$\Delta$ exchanges are negligible. For parameter set B, 
however, the one- and two-$\Delta$ exchanges are sizeable. 
Especially for the Pearce form factor, 
the cross section is enhanced by almost a factor of two when the $\Delta$-exchange 
contributions are added. On the other hand, the overall magnitude of the cross 
sections is, in general, much smaller for set~B than for set~A, 
indicating that the increase due to the $\Delta$-exchange contribution for set B 
is by far not sufficient to compensate the reduction of the $N$-exchange contribution. 
This means that the relative importance of annihilation via 
$\Delta$ exchange as compared to $N$ exchange is much smaller in the 
reaction \mbox{$p\bar{p}\rightarrow \pi^+\pi^-\pi^0$} than it is for 
\mbox{$p\bar{p}\rightarrow \pi^+\pi^-$}. One reason for this is that the number 
of possible charge combinations of the exchanged baryons is larger for 
amplitudes involving $\Delta$'s and
the interferences between their contributions tend to be destructive. The differences
between the results obtained with different form factors are also easily understood
qualitatively. For the Pearce type, since the
factors are in effect associated with propagators, there is one more factor for the
three-pion annihilation amplitude, compared to the two-pion one. For the other types,
however, the factors are associated with internal vertex legs and therefore 
two more factors appear in the three-pion amplitude. 
This makes the Pearce form factor to yield less suppression for annihilation
into three pions. When two baryons are exchanged, each one of them is not as 
far off shell as when only one is exchanged. Therefore, because of its stronger 
variation with $p^2$, the Gaussian form cuts off less strongly 
for three-pion annihilation than the monopole form factor.

The total experimental
\mbox{$p\bar{p} \rightarrow \pi^+\pi^-\pi^0$} cross section ranges from around 7~mb
to 3~mb for energies from 65 MeV to 220 MeV~\cite{Sai}.
In comparison to these values, the calculated uncorrelated cross section 
varies from mere insignificance 
(for monopole and Gaussian form factors with parameter set B), 
to about 10\% of the measured total cross section for the 
Gaussian form factor with parameter set~A. The relevance of the contribution
from annihilation into three uncorrelated pions might be best seen from 
Fig.~\ref{fig8},
where our results for the three uncorrelated pions are added to Mull's
results~\cite{Mull95} for annihilation into two-meson channels 
(\mbox{$\rho\pi$}, \mbox{$f_o\pi$} and \mbox{$f_2\pi$}),
weighted with the percentages of decay of the heavy mesons into two pions and
the isospin factors, and compared with the total experimental annihilation
cross section.

\subsubsection{Branching ratios}
A more detailed comparison can be made by considering the experimental information from a
specific antiprotonium initial state. Annihilation of antiprotonic hydrogen atoms into
\mbox{$\pi^+\pi^-\pi^0$} has been studied by stopping antiprotons from LEAR in hydrogen gas by the
ASTERIX Collaboration~\cite{ASTERIX}. All \mbox{$N\bar{N}$} initial states for $S$ and $P$ waves
which may decay into \mbox{$\pi^+\pi^-\pi^0$} have been considered\footnote{The $G$-parity of the
initial antiprotonium state is \mbox{$G = (-1)^{L+S+I}$}, 
and for a three-pion final state \mbox{$G = -1$};
this determines the isospin quantum number $I$ for each allowed initial state. 
The \mbox{$^3P_0$}
initial state is forbidden by angular momentum and parity conservation.},
namely \mbox{$^3S_1 \: (I = 0)$}, \mbox{$^1S_0 \: (I=1)$}, \mbox{$^1P_1 \: (I = 0)$},
\mbox{$^3P_1 \: (I=1)$} and \mbox{$^3P_2 \: (I=1)$}. The phenomenological analysis has been
made in terms of resonant amplitudes and a non-resonant (phase-space) background.
The contributions
from the \mbox{$\rho\pi$} channel have already been compared with the results of the J\"ulich
model~\cite{Mullrhopi}. The phase-space component has also been compared with the uncorrelated
three-pion contribution in the first, simplified version of our model, in which
$\Delta$ exchange was not considered explicitly. Here we use the present 
model to compare our results with the experimental values.

As before~\cite{us97,Mullrhopi}, we assume that all the spin states of a given orbital angular
momentum in protonium are populated with about the same probability and identify the relative
branching ratios for decay from a given atomic state to the ratios between the contributions
from the corresponding partial waves to the annihilation cross section at low energy. Although
the cross sections themselves vary rapidly with energy, these ratios are
almost constant for small laboratory energies (a few MeV). Thus, like in our previous
work, we use cross-section ratios at 5 MeV.

Table~\ref{table4} shows our results for the ratios of branching ratios for \mbox{$N\bar{N}
\rightarrow \pi^+\pi^-\pi^0$} for various types of form factors and parameter sets, compared to
the experimental values extracted from the phenomenological analysis of Ref.~\cite{ASTERIX}.
An estimate based on ratios between numbers (2J+1) of available states is given as well.
  
Not surprisingly, for parameter set~A, where $N$ exchange is dominant,
 the results are very similar to our
previous ones~\cite{us97}, where $\Delta$ exchange had not been considered at all. 
The theoretical \mbox{$^3S_1/\mbox{}^1S_0$}
ratios depend only weakly on the dynamics. The predictions are close to
the value 3, which is expected on the basis of simple state counting, 
though they tend to be, in general, somewhat larger than this phenomenological 
value, especially for the Pearce form factor. 
The theoretical \mbox{$^3P_1/\mbox{}^1P_1$} ratios show some
dependence on the spin and isospin dynamics  and are smaller than the results of the  
phenomenological analysis by a factor of about 2. 
For the \mbox{$^3P_2/ \mbox{}^1P_1$} ratio, the theoretical results again depend weakly on the
dynamics; however, they also fall short of the experimental evidence by a factor of 2 to 3.

For parameter set B, where $\Delta$ exchange is also relevant, 
the \mbox{$^3S_1/ \mbox{}^1S_0$} ratios show the same general trends
as for set~A, although the result for the Pearce form factor is now considerably larger 
than the experimental value. The  \mbox{$^3P_1/ \mbox{}^1P_1$} and 
\mbox{$\mbox{}^3P_2/ \mbox{}^1P_1$} ratios, on the other hand, are spectacularly larger than for set
A and now overshoot the experimental data by far. This effect originates from the
fact that, with our choices of vertices and propagators, exchanges involving $\Delta$'s make 
a very small contribution to the \mbox{$^1P_1$} partial wave. Consequently, if form factors 
are chosen such as to suppress the \mbox{$NN$}-exchange contribution (as is the case for set B),
the total contribution of that partial wave to the cross section is small and the
\mbox{$^3P_1/ \mbox{}^1P_1$} and \mbox{$^3P_2/ \mbox{}^1P_1$} ratios
are large. The fact that the experimental values for these ratios lie between our results
for sets A and B is in accordance with our objective of these sets as extremes
between which realistic values for the cutoffs are confined.

\subsection{Annihilation into neutral pions}
Some experimental results for \mbox{$p\bar{p}$} annihilation into neutral pions have
been published already a long time ago~\cite{Dulude}. But only very recently, 
for the first time, data at beam momenta below 1 GeV/c have been
made available by the Crystal Barrel Collaboration for annihilation 
into \mbox{$2\pi^0$'s}~\cite{Anisovich1999A,Anisovich1999B}. 
Specifically, their measurement at a laboratory momentum of 600 MeV/c is still
within the limit of validity of our model and, therefore, it is possible to compare 
our results with those data, as we shall do in this section. We
emphasize that all our parameter sets were fixed by the annihilation into two 
charged pions, as discussed above. 
Thus, the results presented here for the reactions 
\mbox{$p\bar{p}\rightarrow 2\pi^0$} as well as 
\mbox{$p\bar{p}\rightarrow 3\pi^0$} are genuine predictions of our model. 

\subsubsection{\mbox{$p\bar{p}\rightarrow 2\pi^0$} annihilation}
For the reaction \mbox{$p\bar{p}\rightarrow 2\pi^0$} there are high-statistics 
data taken at LEAR in the momentum range 600-1940 MeV/c
~\cite{Anisovich1999A}~\footnote{
These data show a systematic disagreement in normalization with
the earlier data~\cite{Dulude}, i.e. the cross sections are a factor of more 
than 2 larger. }. 
In Fig.~\ref{fig9}, we compare the differential cross sections predicted by
our model with the data of Ref~\cite{Anisovich1999A} at the laboratory 
momentum \mbox{$p_{lab} = 600$~MeV/c}. 
(The next higher energy measured, \mbox{$p_{lab} = 900$~MeV/c}, is already beyond 
the valitidy range of our model and therefore we refrain from 
comparing our model with those data.)
Note that the normalization is such that the area under the curve 
times \mbox{$2\pi$} gives the corresponding total annihilation cross section. 
 
It can be seen that the predictions for the differential cross section at small angles
 are reasonable for parameter set~A, i.e. for the models
where nucleon exchange is the dominant annihilation process. 
However, for parameter set B, which corresponds to models with a large
contribution from $\Delta$ exchange, we observe a serious disagreement with the data. 
Obviously the $\Delta$ exchange tends to bend down the differential cross section 
at small angles, something which we also noticed for the charged case. 
For angles near $90^\circ$ the Pearce form factor yields a contribution too 
high for both parameter sets whereas all other models are in rough agreement with 
the data. 

Predictions at a lower laboratory momentum, \mbox{$p_{lab} = 360$~MeV/c}, are also
shown in Fig.~\ref{fig9} though at this energy there are presently no data
available. 

The experimental total cross section for \mbox{$p\bar{p}\rightarrow \pi^0\pi^0$}
given in Ref.~\cite{Anisovich1999A} was obtained by integrating only over
the limited range \mbox{$\cos (\theta)$ = 0} to 0.85. Thus, we do the same for obtaining  
the theoretical cross sections at \mbox{$p_{lab}= 600$~MeV/c}, which are compiled
in Table~\ref{table5}. The model predictions range from values close to the
experimental ones, to values that are too large by a factor of 2 to 3. 
The contributions from the individual annihilation mechanisms ($N$ and $\Delta$
exchange, respectively) are listed in Table~\ref{table6} for all considered
parameter sets. Not unexpectedly, the general features are very similar to those
found for the annihilation into two charged pions. $N$ exchange dominates in
case of parameter set~A, while $\Delta$ exchange provides the larger contribution 
for the parameter set B. The interference between $N$ and $\Delta$ exchanges is
generally constructive.

\subsubsection{$p\bar{p}\rightarrow 3\pi^0$ annihilation }
Our results for the cross section for 
\mbox{$p\bar{p}$} annihilation into three uncorrelated neutral pions at 
\mbox{$p_{lab} = 600$~MeV/c} are compiled in Table~\ref{table7}. 
The values range from \mbox{2.2 $\mu b$} (for parameter set B with monopole form
factor) to \mbox{37.8 $\mu b$} (for parameter set~A with Gaussian form factor).
The corresponding experimental value for the total annihilation cross section 
(which includes contributions from three uncorrelated pions as well as from 
two-meson annihilation channels that finally decay into three pions) 
is \mbox{$356\pm 18$ $\mu b$}~\cite{Anisovich1999452}. Thus, like already for the
case of charged pions, it turns out that the annihilation
into three uncorrelated $\pi^0$'s could amount to up to $10\%$ of the total 
\mbox{$3\pi^0$} annihilation cross section. This suggests that such a 
``non-resonant background'' contribution should perhaps be included in detailed
phenomenological analyses aimed at identifying new hadronic 
states~\cite{Amsler,Abele,Anisovich1999452}.

The contributions from the individual annihilation mechanisms can be seen 
in Table~\ref{table8}. As for the charged case, it turns out that annihilation
via \mbox{$N\Delta$} as well as via double-$\Delta$ exchange is negligible for 
parameter set~A.  For parameter set B, \mbox{$NN$} and \mbox{$N\Delta$} exchanges yield 
similar contributions, 
while \mbox{$\Delta\Delta$} exchange remains practically negligible.
However, the overall size of the cross-section is again much less than for set~A.
The comparison made in Sec.~\ref{sec:3 charged cc} between the various types of form 
factors applies to the present case also.
    
\section{SUMMARY}

In this paper we have presented results of a study of
proton-antiproton annihilation into two and three pions  
in a baryon-exchange model. Specifically, we have taken into
account contributions from $N$ exchange as well as from $\Delta$ 
exchange consistently in the two- and three-pion annihilation channels. 
The free parameters of our model, the cutoff masses in the form factors 
at the \mbox{$NN\pi$} and \mbox{$N\Delta\pi$} vertices, were determined by the 
available experimental data on \mbox{$p\bar p \rightarrow \pi^+\pi^-$}. 
Thus, the results for the other annihilation channels considered,
\mbox{$p\bar p \rightarrow \pi^0\pi^0$} as well as  
\mbox{$p\bar p \rightarrow \pi^+\pi^-\pi^0$} and
\mbox{$p\bar p \rightarrow \pi^0\pi^0\pi^0$}, can be regarded as genuine
predictions of the model.  

The main aim of our work was to study 
and discuss the relative importance of \mbox{$NN$}, \mbox{$N\Delta$} and 
\mbox{$\Delta\Delta$} exchanges for annihilation into three pions. 
For that purpose we prepared two sets of models which describe the
cross sections of the reaction \mbox{$p\bar p \rightarrow \pi^+\pi^-$} with
comparable quality but have dominant contributions from either $N$ or
$\Delta$ exchange. 
In addition we employed three different analytical forms for the 
vertex form factors in order to explore the sensitivity of our results 
to these ingredients of our model. 

It turned out that the contributions from annihilation diagrams
involving the $\Delta$ isobar are, in general, much less important for
the three-pion channel than they are for annihilation into two pions.
Specifically, for those models where the $N$ exchange dominates the
two pion decay we found the $\Delta$-exchange contributions to the
three pion decay to be completely negligible. Even in those models
where $\Delta$ exchange plays a major role in the two-pion channel, 
there is only a moderate effect from the $\Delta$ exchange in the three-pion 
channel. As a consequence, the total annihilation cross section
into three uncorrelated pions is usually significantly larger if we 
assume that $N$ exchange dominates the two-pion decay. 

Not unexpectedly, the actual magnitude of the total three-pion 
annihilation cross section depends to a certain extent on the choice for 
the vertex form factors. This dependence is strongly reduced by requiring
consistency between the treatment of the two- and three-pion decay
channels, but ultimately cannot be avoided because it is already inherent
in the specific functions used for the analytic forms of the form factor.  
With due concession for these uncertainties, 
our calculations show that the contributions of 
annihilation into three uncorrelated pions to the total three-pion
annihilation cross section are by no means negligible. Indeed they might
provide up to 10\% of the total cross section for 
\mbox{$p\bar{p}\rightarrow \pi^+\pi^-\pi^0$} as well as for
\mbox{$p\bar{p}\rightarrow 3\pi^0$}.

The results obtained in this work can also be invoked to assess the general
viability of the baryon-exchange model of nucleon-antinucleon annihilation. 
Although the two-pion and uncorrelated three-pion final states make only rather
tiny contributions to the total annihilation process, the relative sizes
of the cross-sections for these two channels can be used as a hint to evaluate
the possibility of describing most if  not all of annihilation in the baryon-exchange
framework. The cross sections for annihilation into two pions are of the order
of a fraction of a millibarn in the energy region we have considered. The cross sections
predicted by the baryon-exchange model for annihilation into three uncorrelated
pions (taking a rough average of the extreme cases considered in this work, see
Table~\ref{table3}) are of the same order of magnitude. This is perhaps surprising
since one might have thought that the need for more vertices and propagators --
therefore more form factors -- would suppress final states of more than two mesons
in such models. Remembering that, according to the calculations of Ref.~\cite{Mull95},
annihilation into two mesons, summed over all meson types, amounts to about 30\% of
the total experimental annihilation cross section, it is tempting to generalize the
results found here for pions to speculate that annihilation into three mesons, 
when summed over all meson types, could easily account for a similar
percentage of the whole process. Since it seems that multi-meson final states are not
strongly suppressed in baryon-exchange models, states with four or more mesons
could then easily account for the remaining of the cross section. Of course, such qualitative
speculations could only be confirmed by detailed calculations, which would constitute a
formidable task, but it can at least be claimed that the possibility of explaining
the bulk of the annihilation process in a baryon-exchange picture is not ruled out.

\section*{Acknowledgments}
We would like to thank V. Mull for valuable discussions and the ``Centro Nacional de 
Supercomputa\c{c}\~ao da Universidade
Federal do Rio Grande do Sul (UFRGS)'', where much of the computational
work was done. Furthermore, we acknowledge communications with 
A.V. Anisovich concerning the $p\bar p\rightarrow \pi^0\pi^0$ annihilation data. 
Financial support for this work was provided by the
international exchange program DLR (Germany; \mbox{BRA W0B 2F}) - 
CNPq (Brazil; \mbox{910133/94-8}), 
and by FAPERGS (Funda\c{c}\~ao de
Amparo \`a Pesquisa do Estado do Rio Grande do Sul).

\newpage

\begin{table}
\caption{Cutoffs (in MeV) for the various form factors used in the calculations. The first
column gives the type of  form factor; the second one identifies the corresponding set. The
headings of the other columns specify the kind of cutoff.}
\begin{tabular}{lccc}
Type & Set & $ \Lambda_N$ & $\Lambda_\Delta$ \\
\hline
Monopole & A & 1600 & 1300 \\
Monopole & B & 1250 & 1550 \\
Gaussian & A  & 1300 & 1100 \\
Gaussian & B  & 1100 & 1320 \\
Pearce & A & 1200 & 980 \\
Pearce & B  & 1050 & 1250 \\
\end{tabular}
\label{table1}
\end{table}

\vskip 2cm

\begin{table}
\caption{ Contributions to the
\mbox{$p \bar{p} \rightarrow \pi^+ \pi^-$} cross section (in $\mu b$),
calculated with different form factors (first column) and parameter sets
(second column). The third  column gives the laboratory kinetic energy (in MeV), 
the other columns different contributions, specified by the exchanged 
baryon(s) in the annihilation amplitude. }
\begin{tabular}{lccccc}
Form Factor & Set & Energy & $N$ & $\Delta$  & $N+\Delta$ \\
\hline
Monopole & A & 65  & 609 & 1.8  & 685 \\
& &  220  & 322 & 1.1  & 332 \\
\hline
Gaussian & A & 65  & 559  & 6.2  & 663  \\
& & 220  & 375  & 3.7  &  419 \\
\hline
Pearce & A & 65  & 571 & 25  & 782  \\
& & 220  & 304 & 15  & 384  \\
\hline
Monopole & B &  65  & 66  & 436  & 741  \\
& & 220  & 40  & 237   & 357   \\
\hline
Gaussian & B &   65  & 62 &  424 &  693 \\
& & 220  & 54 & 260  & 414 \\
\hline
Pearce & B  &  65  & 182  & 353  & 847  \\
& & 220  & 104  & 189  & 404  
\end{tabular}
\label{table2}
\end{table}

\vskip 2cm

\begin{table}
\caption{Uncorrelated-channel contribution to the
\mbox{$p \bar{p} \rightarrow \pi^+ \pi^-  \pi^0$} cross section (in $\mu b$),
calculated with different form factors (first colmun) and parameter sets
(second column). The third column gives the laboratory kinetic energy 
(in MeV), the other columns list the various contributions.}
\begin{tabular}{lcccccc}
Form Factor & Set & Energy & $NN$ & $N\Delta$  & $\Delta\Delta$ & all \\
\hline
Monopole & A & 65  & 309 & $2.4 \, 10^{-2}$ & $5.7 \, 10^{-5}$ & 310 \\
& &  220  & 200 & $1.7 \, 10^{-2}$ & $4.4 \, 10^{-5}$  & 200 \\
\hline
Gaussian & A & 65  &760  & $3.5 \, 10^{-2}$ & $1.3 \, 10^{-2}$  & 760  \\
& & 220  & 535  & $4.3 \, 10^{-1}$ & $1.4 \, 10^{-2}$  &  535 \\
\hline
Pearce & A & 65  & 603 & 3.2  & $2.4 \, 10^{-1}$ \  & 620  \\
& & 220  & 404 & 2.8  & $1.9 \, 10^{-1}$  & 403  \\
\hline
Monopole & B &  65  & 13  & 3.8 & 1.7 & 20  \\
& & 220  & 8  & 2.5  & 1.2  & 12   \\
\hline
Gaussian & B &   65  & 67 & 12 & 9.4  &  84 \\
& & 220  & 52 & 14 & 8.9  & 75 \\
\hline
Pearce & B  &  65  & 138  & 22 & 46  & 242  \\
& & 220  & 96  & 19 & 34  & 158  
\end{tabular}
\label{table3}
\end{table}

\begin{table}
\caption{Ratios of branching ratios for 
\mbox{$p \bar{p} \rightarrow \pi^+ \pi^- \pi^0$}.
The experimental values are taken from the phenomenological analysis of
Ref.~\protect\cite{ASTERIX}.
The theoretical results are obtained as relative cross sections at
\mbox{$E_{lab}= 5$~MeV} for different form factors and parameter sets.}
\begin{tabular}{llcddd}
& Form factor & Set & $^3S_1/^1S_0$ & $^3P_1/^1P_1$ & $^3P_2/^1P_1$ \\
\hline
& Monopole &   & 3.3 & 1.7  & 1.4  \\
Model & Gaussian & A  & 3.2 & 1.5  & 1.1 \\
& Pearce &   & 3.9 & 1.7  & 1.8 \\
\hline
& Monopole &   &3.2  & 24.3  & 7.9 \\
Model & Gaussian & B  & 2.9 & 7.1  & 5.6 \\
& Pearce &   & 6.3 & 34.9  & 13.6 \\
\hline
Experiment  &  &   & 1.9 & 2.9  & 3.8 \\
\hline
State counting & &  & 3.0 & 1.0 & 1.67\\
\end{tabular}
\label{table4}
\end{table}

\begin{table}
\caption{Cross section for \mbox{$p\bar{p} \rightarrow \pi^0\pi^0$} 
at 600 MeV/c, integrated over angles from \mbox{$\cos (\theta)$ = 0} to 0.85. 
The data is from Ref.~\protect\cite{Anisovich1999A}.}
\begin{tabular}{lcd} 
Type & Set & cross section ($\mu b$) \\
\hline
         Monopole & A & 53.8   \\
         Monopole & B & 106.9  \\
         Gaussian & A  &  80.7   \\ 
         Gaussian & B  & 119.8   \\
         Pearce & A &  100.4    \\
         Pearce & B  &  126.5   \\ 
\hline
         Experiment & & 54.4    \\
\end{tabular}
\label{table5}
\end{table}

\begin{table}
\caption{The same as Table~\ref{table2} for the \mbox{$p \bar{p} \rightarrow 
\pi^0  \pi^0$} annihilation.}
\begin{tabular}{lccddd}
Form Factor & Set & Energy & $N$ & $\Delta$ & $N+\Delta$ \\
\hline
Monopole   & A & 65 & 173  & 8.1 $10^{-1}$ & 195 \\
           & &  220  & 62  & 4.5 $10^{-1}$ & 70 \\
\hline
Gaussian & A &  65  & 232  & 3  &  283 \\
& &            220  & 92  & 1.7  & 112  \\
\hline
Pearce & A &      65  & 257  & 11.9    & 359   \\
& &              220  & 87 &  6.7    & 127  \\
\hline
Monopole & B & 65  & 21   & 197   & 328   \\
      & &     220  & 8  &  100  & 146   \\
\hline
Gaussian & B &  65  & 20 & 204 &  323\\
& &            220  & 13  & 120   & 175 \\
\hline
Pearce & B  &   65  & 81 &  167   & 397   \\
& &            220  & 29 & 83   & 167  \\
\end{tabular}
\label{table6}
\end{table}

\vskip 2cm

\begin{table}
\caption{Uncorrelated contribution to the cross section for 
\mbox{$p\bar{p} \rightarrow \pi^0\pi^0\pi^0$} at 
\mbox{$p_{lab} = 600$ MeV/c}, calculated
with different form factors and parameter sets. Note that the experimental
total annihilation cross section (which includes the uncorrelated as well
as correlated contributions) is 
\mbox{356 $\pm$ 18 $\mu b$}~\protect\cite{Anisovich1999452}. 
}
\begin{tabular}{lcd} 
Type & Set & $ \sigma (\mu b)$ \\
\hline
              Monopole&  & 13.5   \\
              Gaussian& A  &   37.8  \\
              Pearce&  &    31.8  \\
\hline
              Monopole&  & 2.2   \\ 
              Gaussian& B  &  7.2  \\
              Pearce&   &   16.0  \\ 
\end{tabular}
\label{table7}
\end{table}

\vskip 2cm

\begin{table}
\caption{The same as Table~\ref{table3} for the \mbox{$p \bar{p} \rightarrow 
\pi^0 \pi^0 \pi^0$} annihilation. All cross sections are given in \mbox{$\mu b$}.}
\begin{tabular}{lcccccc}
Form Factor & Set & Energy & $NN$ & $N\Delta$ & $\Delta\Delta$ & all \\
\hline                   
 Monopole & A & 65 & $24 $ & $9.6 \, 10^{-3}$  & $8.4 \, 10^{-7}$ & $25 $ \\
 & &           220  & $11 $ & $4.3 \, 10^{-3}$ & $4.9 \, 10^{-7}$ & $12 $ \\
\hline
Gaussian & A & 65 & $73 $  & $0.097 $   & $3.3 \, 10^{-4}$& $74 $  \\
& &           220  & $32 $  & $0.060 $& $2.6 \, 10^{-4}$  & $31 $ \\
\hline
Pearce & A & 65 & $51 $ & $1.2 $ & $3.6 \, 10^{-3}$   & $60 $  \\
& &         220  & $23 $ & $0.53 $ & $2.4 \, 10^{-3}$  & $27 $  \\
\hline
Monopole & B & 65  & $ 1.0 $   & $1.6 $& $0.025 $  & $4.2 $  \\
& &            220  & $0.46 $  & $0.75$ & $0.014 $   & $1.9 $   \\
\hline
Gaussian & B & 65  & $7.8 $ &  $4.3 $& $0.22 $ &  $14 $ \\
& &            220  & $3.6 $ & $2.2 $& $0.15 $  & $6.4 $ \\
\hline
Pearce & B  & 65  & $12 $  & $9.7 $& $0.53 $  & $31 $  \\
& &          220  & $5.6 $  & $4.6 $ & $0.32 $  & $14 $  \\
\end{tabular}
\label{table8}
\end{table}

\newpage

\begin{figure}
\caption{Born transition amplitude for \mbox{$N \bar{N} \rightarrow 2\pi$}.
}
\label{fig1}
\end{figure}

\begin{figure}
\caption{Born transition amplitude for \mbox{$N \bar{N} \rightarrow 3\pi$}.
}
\label{fig2}
\end{figure}

\begin{figure}
\caption{Elastic part of the \mbox{$N \bar{N}$} interaction.
}
\label{fig3}
\end{figure}

\begin{figure}
\caption{Microscopic (a) and phenomenological (b) annihilation part of the  
\mbox{$N \bar{N}$} interaction.
}
\label{fig4}
\end{figure}

\begin{figure}
\caption{Transition potential for \mbox{$N \bar{N} \rightarrow 2 \,{\rm mesons}$}.
}
\label{fig5}
\end{figure}

\begin{figure}
\caption{\mbox{$p\bar{p} \rightarrow \pi^+\pi^-$} cross sections.
The solid lines, long dashed lines and short dashed lines correspond
to our results with the monopole, Gaussian and Pearce form factors.
At left (right), results with parameter set~A~(B).
The data are from Ref.~\protect\cite{Sai,Sugimoto,Bardin}.
}
\label{fig6}
\end{figure}

\begin{figure}
\caption{\mbox{$p\bar{p} \rightarrow \pi^+\pi^-$} differential cross sections 
for \mbox{$p_{lab} = 360$ MeV/c} and for \mbox{$p_{lab} = 679$ MeV/c}. 
Same description of the curves as in Fig.~\ref{fig6}. Experimental data are 
taken from Ref.~\protect\cite{Hasan}.
}
\label{fig7}
\end{figure}

\begin{figure}
\caption{\mbox{$p\bar{p} \rightarrow \pi^+\pi^-\pi^0$} cross section as a function
of the incident momentum. The sums of two-meson channels \protect\cite{Mull95}
plus our results for the three  uncorrelated-pion channel with the various types
of form factors and parameter sets stated in Table~\ref{table1} lie inside the 
shadowed area. Experimental data are taken from Ref.~\protect\cite{Sai}.
}
\label{fig8}
\end{figure}

\begin{figure}
\caption{\mbox{$p\bar{p}\rightarrow \pi^0\pi^0$} differential cross sections 
for \mbox{$p_{lab} = 360$ MeV/c} and for  \mbox{$p_{lab} = 600$ MeV/c}. 
The same description of the curves as in Fig.~\ref{fig6}.
Experimental data are taken from Ref.~\protect\cite{Anisovich1999A}. }
\label{fig9}
\end{figure}


\newpage

\begin{figure}[hbct]
\includegraphics{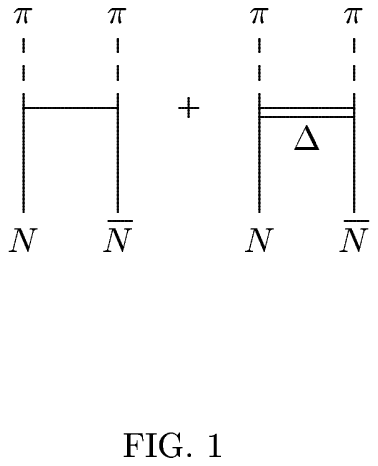}
\end{figure}
\vspace{5cm}
\center{  }

\newpage

\begin{figure}[hbct]
\includegraphics{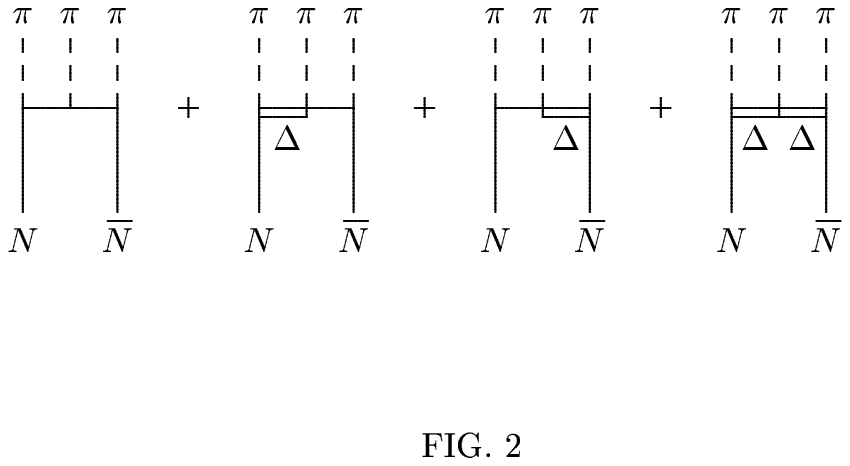}
\end{figure}
\vspace{5cm}
\center{  }

\newpage

\begin{figure}[hbct]
\includegraphics{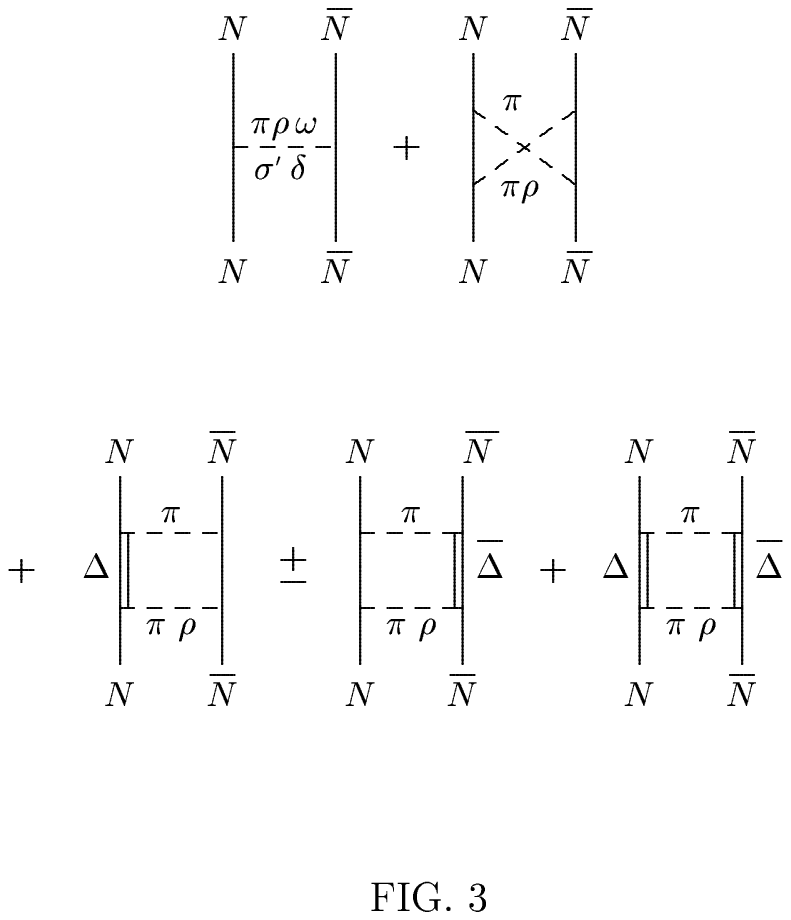}
\end{figure}
\vspace{5cm}
\center{  }

\newpage

\begin{figure}[hbct]
\includegraphics{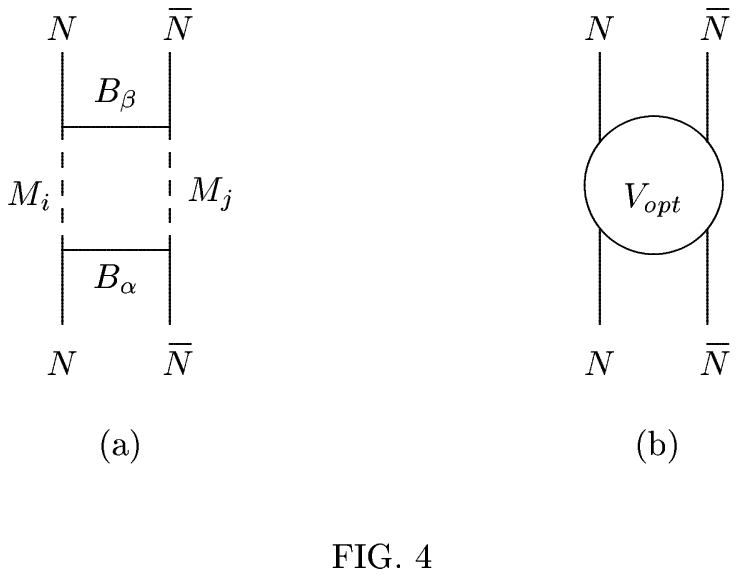}
\end{figure}
\vspace{5cm}
\center{  }

\newpage

\begin{figure}[hbct]
\includegraphics{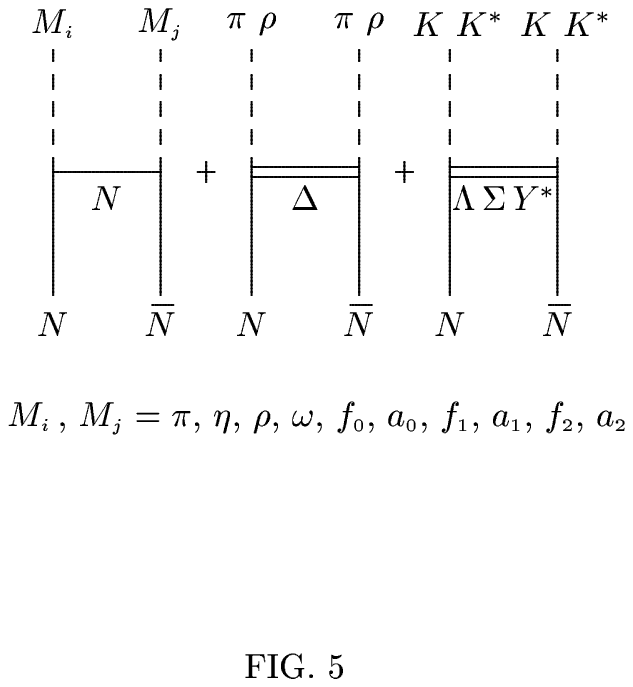}
\end{figure}
\vspace{5cm}
\center{  }

\newpage

\begin{figure}[hbct]
\includegraphics{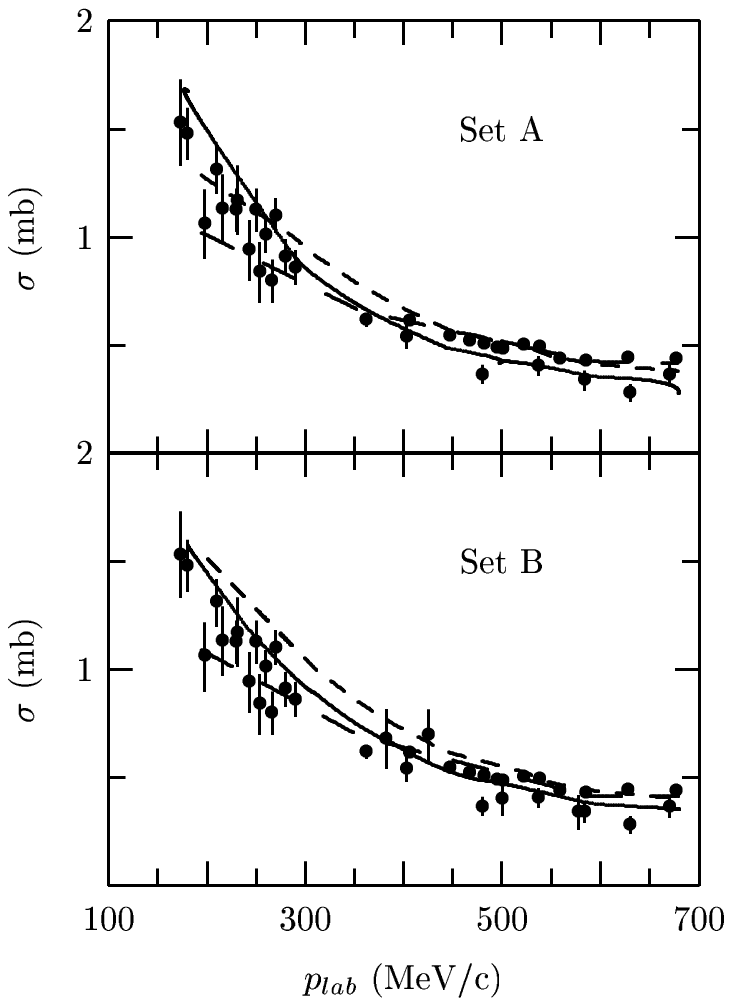}
\end{figure}

\vspace{5cm}
\center{FIG. 6}

\newpage
\begin{figure}[hbct]
\includegraphics{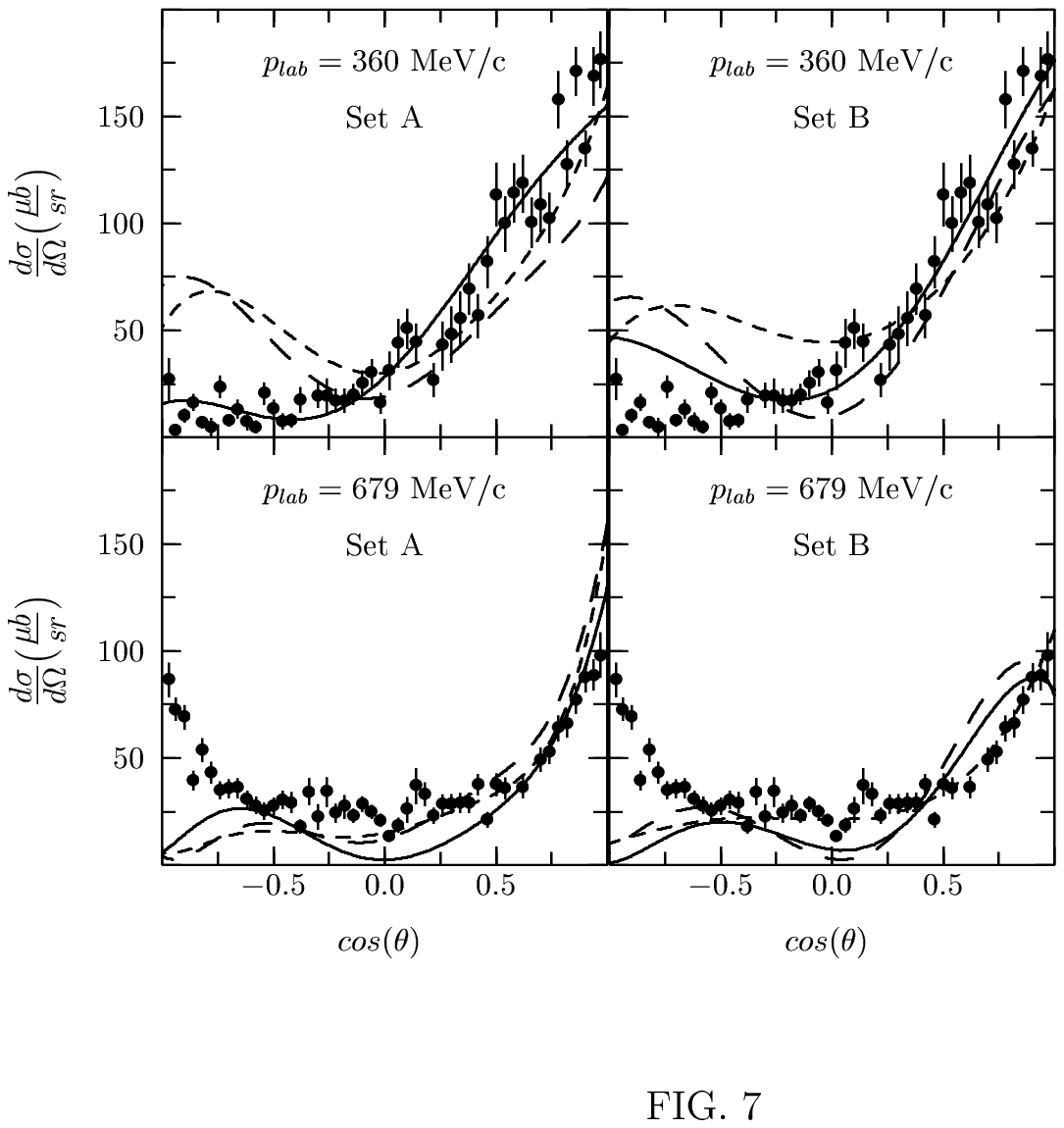}
\end{figure}

\vspace{5cm}
\center{FIG. 7}

\newpage
\begin{figure}[hbct]
\includegraphics{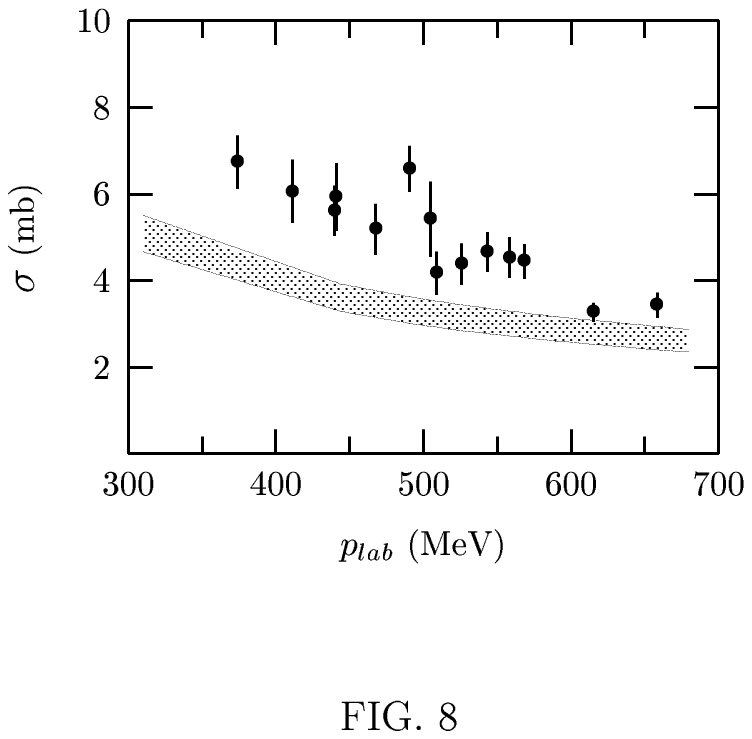}
\end{figure}

\vspace{5cm}
\center{FIG. 8}

\newpage
\begin{figure}[hbct]
\includegraphics{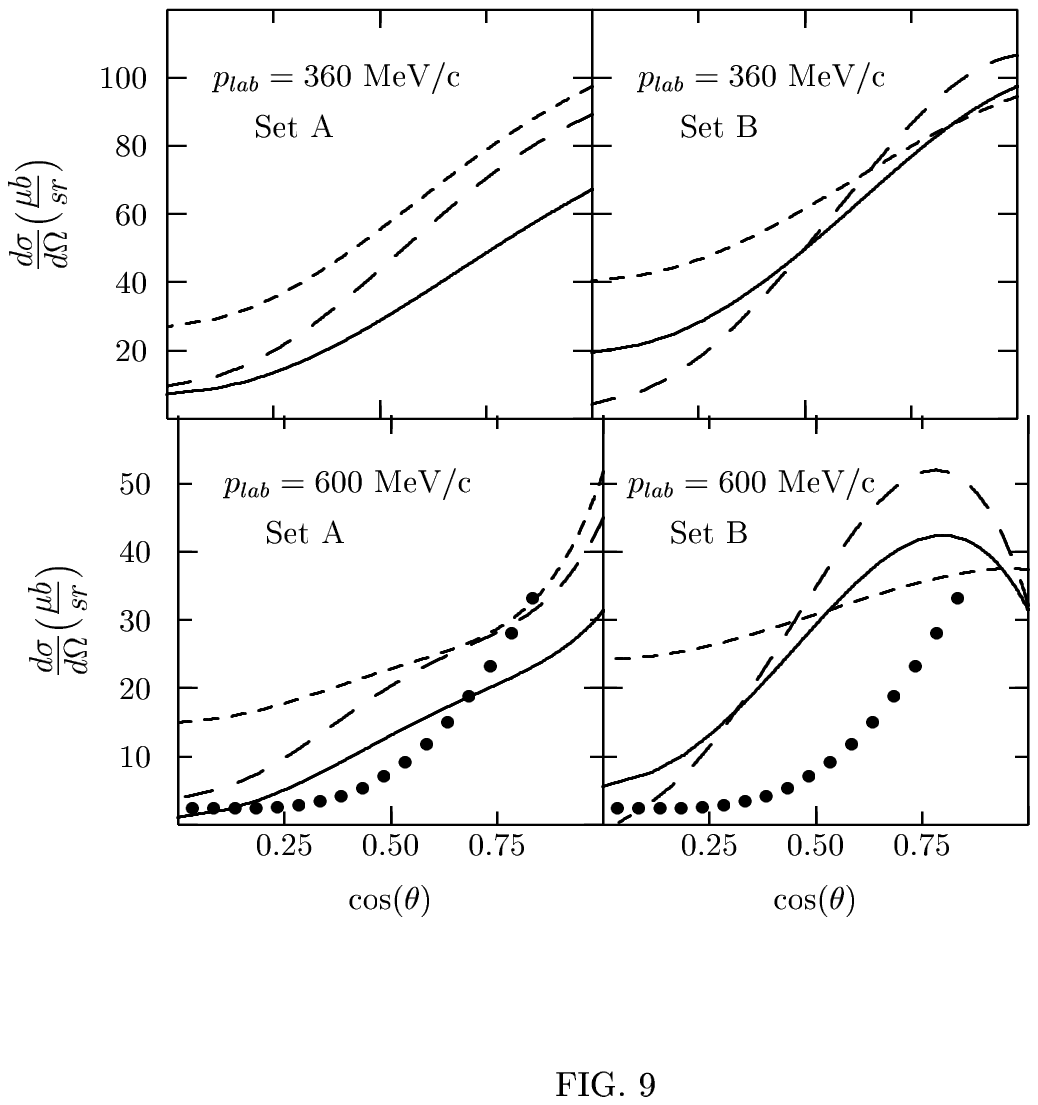}
\end{figure}

\vspace{5cm}
\center{FIG. 9}

\end{document}